\documentclass[prd,twocolumn]{revtex4}
\usepackage{graphicx}
\usepackage{dcolumn}
\usepackage{bm}
\begin{document}
\title{ Estimate of resources required for a meaningful reform of education }
\author{ Stanis{\l}aw D. G{\l}azek }
\date{v1 31 Jan 2007; 
v2 16 Oct 2010;
v3  1 Jul 2012;
v4 23 Nov 2012}
\begin{abstract}
A simple estimate in terms of currency units 
shows that a meaningful educational reform 
process can be launched and sustained over 
many generations of teachers with support 
of parents of students. In the estimate, the 
steady inflow of resources from parents 
provides support for advanced studies by 
teachers. Not to waste the resources on 
spurious activities, the estimated inflow 
proceeds directly from the parents as clients 
to the providers of required reform program. 
The providers are the experts in various 
disciplines who excel in helping teachers 
become great. Their services to teachers are 
ultimately assessed by parents on the basis 
of changes in behavior of children. The resulting 
reform program grows slowly from small seeds. 
The running cost of the reform process to
parents appears surprisingly low while its 
development leads to the desired changes over 
time.
\end{abstract} 
\maketitle

\section{ Introduction }
\label{I}

Ongoing discussions and practices of educational
reform movements reflect a variety of ideas and
many conflicting interests. The net result is 
not satisfactory. Students do not learn what 
they are said to be taught. The worst thing is, 
however, that the acute problems with facilitation 
of learning continue to grow at all levels of 
society and lead to social disparities and unrest 
instead of being gradually solved as they appear.

Among many, the key reason for this result is 
that reform efforts are typically {\it not} 
centered on the people who learn and teach. 
Most reform efforts do {\it not} take seriously 
how students and teachers see themselves and how 
they imagine their own future. Instead, typical 
reform programs appear tuned predominantly to 
what the reformers are interested in achieving, 
irrespective of the states of mind of students 
and teachers. Thus, the students and teachers do 
not have any deeply vested interest in engaging 
in typical reforms. The reforms are not ``theirs'' 
and it is natural for them not to engage every 
year in another ``reform'' administered from 
above. It is more economic for them personally 
to ignore such fads. 

Consequently, the changes that reformers promise
when they advertise their reform projects do not 
occur. Instead, the vagaries of reform continue and 
fail to create a qualitatively new system that 
would guarantee {\it by design} that the majority 
of alumni are competent learners with a sizable 
probability of achieving satisfaction in life. 
The educational system is stuck in such vagaries 
of reform for a long time, at least an order of 
magnitude longer than the average duration of 
employment of an average teacher. To get out of 
this cycle of reform failure, the system needs a 
sufficiently strong force of change to start 
working in it and never stop. 

The leading hypothesis of the estimate described
here is that if teachers take initiative in 
their hands and become co-authors of the reform 
program with support of parents, the program has 
a chance to develop over sufficiently many 
generations for the desired, qualitative change 
to eventually happen. The estimate suggests that 
parental financing of the reasonable reform 
program is realistic.

\subsection{ Foundation }
\label{F}

In the estimate, the process of preparation 
that teachers can consciously engage in for 
eventually becoming great is based on 
deliberate practice of learning in the context 
of life, work, and disciplines in which teachers
specialize. Such successful preparation is an 
example of learning that we call productive. 
Productive learning by teachers leads them to 
the concept of productive teaching. This means 
that students learn productively in the contexts 
of their lives and take advantage of and develop 
their strengths as they study relevant disciplines 
coached by their teachers.

On the basis of an assumption that a new educational 
system will develop as a result of productive teaching, 
it is assumed in the estimate that the new system will
be gradually formed in a process of one generation of 
teachers learning from another. This process is seen
as analogous to the long-term natural evolution of 
species, except that humans are capable of adding some 
conscious action to the unconscious ways of Nature. 
Thus, {\it the main idea underlying the estimate is  
that the natural learning mechanisms in parents, 
students, teachers, and experts of all disciplines 
can be engaged in building the new system.} 

Since productive learning occurs when one 
generation successfully learns from another, 
the ultimate providers of resources for 
teaching new generations are their parents. 
{\it Reform program will be steadily supported 
by parents if and only if it facilitates 
productive learning by children. Such facilitation 
requires that teachers 1) consciously learn 
according to their wants, needs and abilities 
coached by successful experts in the disciplines 
they specialize in and 2) become independent 
developers of their own expertise on productive 
teaching as the signature of their profession. 
The role of parents is to support teachers in 
creating, sustaining, and participating in 
both these processes.}

This note estimates the order of magnitude of 
the costs of a reform process that parents 
would have to cover on regular basis. For 
productive teaching to become a commonplace, 
it is necessary that conscious facilitation 
of productive learning by teachers occurs 
over sufficiently many generations to become 
a systemic principle: a feature and a paradigm. 
Certainly more than two generations must be 
involved for more than just one generation of 
teachers to learn from another what it takes 
to become a great teacher. 

The condition of more than two generations 
being involved implies a period on the order 
of at least half a century. Such a time scale 
does not lie within the time horizons considered 
in typical reform programs. 

The estimate of required resources is provided 
here in terms of currency units. The resulting 
numbers suggest that launching and sustaining a 
meaningful reform process is not as unrealistic 
as one might expect on the basis of experience 
with failure of reforms that are not systemically 
supported by more than one generation of parents 
and thus do not involve teachers as professionals 
who systematically improve their performance 
generation after generation.

\subsection{ Key problem and the way around it }
\label{KeyProblem}

Among many problems that need to be solved 
in order to develop a meaningful reform as 
indicated above, the central one is the need 
for a clear definition of what is meant by 
productive learning. No such clear definition 
is explicitly shared by all members of society. 
In particular, no such concept is currently 
shared by the reformers, teachers, students, 
and parents; certainly not to a degree of 
clarity that could even remotely compare with, 
for example, the degree of clarity of the 
concept of flying on an airplane that is 
shared among the aviation industry and their 
clients. 

It is thus imperative that the estimate described 
in this note appears sufficiently motivating for 
at least some participants in a current system so
that they can start contributing to a buildup of 
a reasonable reform process even if the society at 
large does not yet clearly define and commonly share 
any overarching concepts of productive learning and 
teaching. So, let the underlying principle be described
in terms of the following four points: 
\begin{enumerate}
\item
The only long-term providers of resources for a 
meaningful educational reform are the parents of 
students; 
\item 
The only way parents can significantly contribute 
to the improvement of educational regularities is 
to provide support for advanced studies by teachers;
\item
Teachers need extensive studies and cooperation 
with scientists, artists and professionals in 
order to become able to help students actively 
prepare for successful functioning in society 
according to their strengths and learning skills;
\item
Current educational systems in general do not 
offer opportunities for the required studies 
and cooperation as fundamental systemic features.
\end{enumerate}

None of these points explains what constitutes 
productive learning and teaching in any 
operational form. Nevertheless, one can assume 
that some parents understand these points
sufficiently, even if only intuitively (see 
below), to support the process of helping 
teachers to become great. These parents could 
become the initial providers of support needed 
for launching the process of meaningful reform. 

If a suitable organization of teachers existed,
all parents could be offered opportunities to 
provide their support for advanced studies by 
teachers. However, only a subgroup of parents 
who are lucky enough to get motivated by teachers
of their children have a chance to act before 
the society commonly agrees on what constitutes 
productive learning, how productive teaching 
serves the purpose, and how large the cumulative 
magnitude of gradual changes will be accomplished 
during the next 100 or 1000 years. Therefore, the 
principles of productive learning and teaching 
that underlie the advanced studies by teachers 
can only gradually be implemented in practice 
based on a gradual increase of parental support. 
Such gradual increase of support may stimulate
building of a new educational system before a 
broad consensus about the meaning of productive 
learning and teaching is achieved.

That the underlying principles may work this way 
around the current lack of shared concepts of 
productive learning and teaching is suggested 
by analogy with the fact that there is no hard 
evidence of any plan on the part of Mother Nature 
to produce mankind step by step in terms of small 
changes that occur generation after generation and 
yet mankind exists and excels among animals as a 
result of evolving from generation to generation 
by small steps. The crux of this note is that one 
can adopt the step by step strategy for reforming 
education.

The idea of reform action that is formulated here 
in terms of the estimate of resources can be 
pondered about in terms of the question ``How?'' 
instead of ``What?'' The word ``How'' refers to 
the method while the word ``What'' refers to the 
content of what is taught. Over time, relatively 
quick changes occur in ``What'' because civilization 
changes relatively quickly on Darwinian scale of 
time. But ``How'' people learn is unlikely to 
significantly change over a millennium. From the 
point of view of the crux of this note, changes in 
technology, such as from a stick on sand to ink on 
paper to computer display to telecommunication, are 
very important but they only concern the ongoing 
change of tools. The teaching process, in contrast, 
continuously creates and changes a living person. 
No chisel and hammer can properly work for this 
purpose in the hands of a school employee who has 
not personally lived through a process of productive 
learning and has no examples to follow in developing
knowledge and skills required in productive teaching. 
Students need teachers who know and understand the 
mechanisms of how a living person truly learns. 
These mechanisms are not changing quickly on 
Darwinian scale of time, irrespective of the 
technological advances.

\subsection{ Reservations }
\label{R}

Whereas this note does not itself refer to a lot 
of empirical evidence, the author has arrived at 
the ideas described here on the basis of studies 
conducted with the participation of students and 
teachers and a lot of discussions with experts. 
This note follows three earlier documents that 
contain lists of some relevant sources. The earlier 
documents can be treated as supporting material 
to this note. The first is the article {\it The 
Living Network of Schools Owned by Teachers and 
Students} from 1998~\cite{Network}. It exhibited 
a great enthusiasm for the idea that understanding 
of one thing leads to understanding of another and 
eventually one can overcome all hurdles. In the 
years since then, the author has lived through 
several periods of renewed enthusiasm and despair 
regarding educational reform. Two outcomes of this 
struggle are Ref.~\cite{ProductiveLearning} from 
2006, and Ref.~\cite{HMOT} from 2008. 

\subsection{ Outline of next sections }
\label{outline}

Section~\ref{sec:dream} describes a common 
problem of funding that planners of educational 
reforms face when they realize that their reform 
plan may not come true. 

On this background, Sec.~\ref{P} discusses some
issues that matter to the idea of parents providing 
support for advanced studies by teachers in 
collaboration with experts. The discussion includes 
the following issues: 
why would teachers want to work with experts, 
Sec.~\ref{WhyTeachersWork};
could parents decide to pay for help of experts,
Sec.~\ref{ParentsCanDecide}; 
could parents currently buy learning for their 
children on regular basis, 
Sec.~\ref{CanParentsBuy};
are they satisfied with the status quo,
Sec.~\ref{Satisfied};
what can parents do about it,
Sec.~\ref{WhatDo};
whom can they pay for help, 
Sec.~\ref{WhomPay};
is learning an element of common curricula,
Sec.~\ref{IsLearningInC};
how much parents can pay,
Sec.~\ref{HowMuch};
how students can contribute,
Sec.~\ref{HowStudents};
how parents can avoid wasting money,
Sec.~\ref{AvoidingWaste};
and, the need for small trials,
Sec.~\ref{SmallTrials}.

Subsequently, Sec.~\ref{sec:calculation} describes 
an approximate calculation of the support that a 
society can naturally provide for development of 
a self-improving and growing network of professional 
teachers. The calculation involves the following 
elements: 
initial assumptions used in the estimate,
Sec.~\ref{assumptions};
the role of informing parents about the project,
Sec.~\ref{informing};
estimate of the funds realistically available 
for advanced studies by one teacher for a year,
Sec.~\ref{costperteacher};
the concept of a Teachers Center, or TC, to which 
the tuition accumulated for a teacher by parents 
is paid,
Sec.~\ref{tuition};
statement of the goal of TC,
Sec.~\ref{TCgoal};
crunching numbers to estimate parental and other 
funds that can be made available for running a TC, 
Sec.~\ref{TCnumbers};
estimate of productivity of TCs,
Sec.~\ref{TCproduct};
estimate of funds per employee of a TC,
Sec.~\ref{perTCemployee};
estimate of the workload of a TC,
Sec.~\ref{TCwork}.

After it is estimated that parents are able
to provide enough resources for a TC to exist,
Sec.~\ref{TCformation} discusses main issues 
concerning formation of a TC. The following
issues are discussed:
the role of a TC director as its founding 
leader and future employee,
Sec.~\ref{d};
after-school programs as laboratories where 
teachers and experts can carry out educational 
experiments with participation of students, 
Sec.~\ref{aflab};
the role of initiative being in the hands of 
teachers,
Sec.~\ref{initiative};
self-selection of founders,
Sec.~\ref{self-selection};
the new role to play by experts in pedagogy 
and andragogy,
Sec.~\ref{PandA};
location and operation of a TC,
Sec.~\ref{TClocation};
summary of where the promise of the future for 
a TC comes from,
Sec.~\ref{promise};
examples of suggestions for the initial action 
by founders of a TC,
Sec.~\ref{firstaction}.

Section~\ref{NTCbrain} speaks briefly about an 
international organization that could perform 
the function of a brain in the network of TCs. 

The concluding Sec.~\ref{sec:c} warns readers 
about serious opposition that both the idea of 
and attempts to create a new, global educational 
system with its own brain is likely to encounter.

\section{ The reformers funding dream }
\label{sec:dream}

Consider the example of Ref. \cite{Network} 
in which a vision of a new educational system 
is sketched freely. The system is described 
as ``powered at all levels equally by the will 
of students to learn and the will of teachers 
to learn and share their expertise with 
students.'' The author wished that such system 
is created. The dream was that all one needs
to do is to communicate with the people who are 
interested in creating the system and, as a 
result of shared understanding of the system 
concept, get to work together. Some of that has 
actually happened in the form of a small-scale 
project. However, the results were different 
than intended. The author had to abandon illusions, 
focus on unlearning, and start thinking from 
scratch about how Mother Nature could actually 
produce humans as the most advanced learning 
creatures on earth and how She continues to secure 
that humans apparently accelerate in learning. 

The small project mentioned above is used here
to illustrate how the problem of a lack of 
sufficient support surprises reformers even
if they are able to convince themselves that 
they know what matters and proceed accordingly. 
Precisely in this spirit, the project was conceived 
in advance to sustain itself and grow thanks to 
the high quality of science education it would 
provide to teachers and thus to their students. 
The quality was to follow from the competence 
of people who carry out the project. An entire 
system was supposed to grow out of the initial 
activity because of its sound rules and noble 
goals. The financial aspects were carefully 
managed and nothing could happen without support 
coming from those who were served by the project. 

The project certainly appeared as aiming at 
improving teaching. But the numbers of 
participating students and teachers were too 
small to secure funding of its growth. The 
instructors could not always volunteer or 
work full-time because they had to support 
their families while the long-term prospects 
of earning a living were tangible only within 
the existing system, i.e., {\it not} within a 
small, new enterprise that could be shuttered 
at any time. It was not clear to the team who 
might be the key providers of support for a 
system of serious learning by teachers, for 
what reasons and purposes. In any case, the 
dream system could not sprout from their 
small seed without committed external help, 
i.e., investment. 

Therefore, the team prepared a proposal to a 
major foundation in which it requested funding 
for a detailed plan of development based on the 
experience accumulated in the pilot project~\cite{CN}. 
Again, the central assumption was that after some 
time the fledgling little system would get on its 
feet and begin stable growth. The funding would 
expire but the system was meant to be designed well 
enough to become capable of sustaining itself by 
providing services that teachers truly need for 
learning how to serve their students effectively. 
It was still not clear where from the teachers 
would obtain support to pay for the services they 
would receive. It was subconsciously assumed that
somehow the existing system would become interested 
in purchasing the project services.

The above scenario is an example of the reformers'
dream that assumes that what the reform effort
intends to achieve in a thought-through fashion
is so good that there is no way the program is 	
going to die once it is given a chance to unfold 
its true potential. The belief behind such thinking 
is analogous to the dream that parents have when 
they tend for their child and imagine that the child 
will succeed in the future if only they provide the 
opportunity. 

Reformers face the fact that their reform programs
will die without funding. Some reformers learn to 
excel in continuously securing and managing new 
funds for continuing the work they believe worth 
continuation. However, obtaining funds is not 
synonymous with productive learning and teaching. 
Neither funding nor inflow of other resources is 
the whole story. The question is why the resources 
flow in and why they are expected to continue to 
flow in. A reform program has a chance for initiating 
a process of qualitative change to a new level of 
educational practice if it recognizes the features 
and benefits of productive learning and teaching 
and if it involves the corresponding systemic 
mechanisms through which the beneficiaries pay for 
functioning of the program.

Such clear mechanisms of support are needed in the
educational change program not only to sustain it 
physically but also to make it attractive and 
promising mentally. People need to see that what 
they can contribute by joining the program may pay 
back in ways not limited to financial income. They 
need to see that they will have a chance to grow as 
persons if they contribute. People want to become -- 
come closer to -- who they want to be and who they 
think they can be. This is why they want to learn. 
One may suspect that their brains subconsciously 
respond to the bio-physical tension accumulated 
somewhere in their tightly folded genetic codes 
that are ready to unfold in suitable conditions 
(no precise scientific evidence for such simple 
interpretation of human condition is known to the 
author). To attract new individuals to a long-term 
educational project, not only the inflow of resources 
must be realistically imaginable but also the goals 
as such must be attractive.

In summary, what ultimately counts when people 
assess the odds of success of ``a great reform idea'' 
and when they decide about joining the effort are 
the outlook for survival and growth of the project 
and the quality of reasons for the outlook to be 
good. Obtaining some form of temporary funding for 
a reform of education is not a key to lasting change. 
Instead, the key appears to be the clear mechanism 
that can secure the project survival and growth over 
long time. Such mechanism is, essentially, the promise 
of the future, the outlook that humans need to engage 
in action. 

The problem of funding of educational change that 
is addressed in this note in terms of an estimate 
concerning required resources can be posed in terms 
of two questions. {\it What will pump energy into a 
reform program in amounts sufficient for its function 
and growth over a long time? Where is the future to 
come from?} 

\section{ Preliminaries }
\label{P}

In order to carry out the calculation, one has 
to name the providers and users of the resources. 
One has to explain why the resources will flow 
in the desired direction, and why the flow will 
be sustained. The stress is put on the condition 
that ``the flow of resources in the desired direction 
will be sustained.'' The calculation should tell 
the prospective participants in a reform project 
to what extent the promise of the future in it 
is realistic and why. The essence of the calculation 
is that it predicts a measurable number, i.e., the 
promise of the future is expressed in terms of a 
number that can be verified by comparison with data 
as the project develops. 

Here is the central hypothesis of the calculation.
{\it The ultimate inflow of energy to the educational 
reform program comes form parents and students.} 

This is meant literally. The parents and students 
are the living organisms that gather energy from the 
environment and they bring it to the reform program 
to sustain its function, growth, and improvement. 

Parents pay in more than one way, and students 
borrow against their future work to pay for 
services they receive. Essentially, in the
estimate described here, parents and students 
enable teachers financially to hire experts who 
can help them improve teaching. The improvement 
in teachers' performance is the mechanism of 
reform in which systemic changes are created 
by teachers on the basis of their increased 
awareness, knowledge, and understanding of the 
value that they bring to the students and their 
parents. They want to bring this value to them 
ever more productively as they themselves learn 
more and thus know and understand better and 
better what they are actually doing. 

The necessary condition for the sustained inflow 
of energy from parents and students to teachers 
for the purpose of hiring experts who qualify for 
helping teachers in advanced studies and improvement 
of teaching is that the inflow proceeds along a 
clear path from clients to the providers. Otherwise, 
there will be wasteful activities growing around 
the program. This is how the parasites live. To 
prevent the growth of unnecessary activities, the 
transparency of the resource flow must be constantly 
protected. For example, the transparency prevents 
excessive bureaucracy and profit seekers from becoming 
a burden on a reform program. Transparency secures 
such prevention because the parasites cannot clearly 
justify their participation in true learning and 
teaching processes; unclear arguments can be 
identified and rejected.

\subsection{ Why teachers would want to work with experts }
\label{WhyTeachersWork}

True teachers are learners themselves. They desire 
to know, understand, and achieve more than they 
already do. They understand that their own growth 
is the basis of their ability to set example and 
thus encourage their students to grow and help 
them to grow. In addition, if they have a chance 
to participate in building a system in which their 
competence as teachers is rewarded, they can be 
expected to want to participate. On the basis of 
social studies one may expect that about half of 
teachers may be inclined and a smaller fraction may 
be eager to participate in educational research and 
development activities.

\subsection{ Parents can decide to pay for help of experts }
\label{ParentsCanDecide}

Most parents think that their children need to learn 
in order to achieve happiness in life. However, not 
all parents are thinking about their children's lives 
in terms of discovery of what one's cup of tea is and 
how to drink it. Typically, parents know their own life 
and they wish their children to do better. Regarding 
teachers, parents remember a teacher or two who were 
helpful to them in school. Once parents are informed 
about how and why and with what results experts can 
help teachers increase their competence, they have a 
chance to compare whom they had and whom their children 
may have as teachers. On this basis, they can decide 
to pay.   

\subsection{ Can parents currently buy learning? }
\label{CanParentsBuy}

The question is whether parents can buy high-quality 
teaching services for their children from typical 
teachers in typical public schools (of all levels). 
Typically, the answer is no, they cannot. Moreover, 
the concept of a public school in the current system 
is based on the assumption that parents do not and 
can not pay directly to the teachers. Instead, they 
pay taxes and the taxes are redistributed by the 
government, only part ending up in the public education 
system. For example, currently about 40\% of local 
income taxes in Poland is at the discretion of a 
local government. In unprivileged areas even up to 
80\% of this amount may have to be spent on schools 
to merely sustain their operation (in Warsaw, Poland 
schools consume about a quarter of the city budget). 

The quality of service to students that is bought 
using the public money is not easily verifiable. 
The teachers' income in public schools does not 
directly depend on the quality of their services. 
It is not clear to what extent the members of a 
government system understand what members of the 
educational system are doing or could be doing 
differently. Parents have no alternative to 
governmental distribution of public funds for 
supporting advanced learning by teachers and 
thus they have no direct way to influence the 
extent of opportunities of their children to 
learn under guidance of the teachers.

\subsection{ Are parents satisfied? }
\label{Satisfied}

Unfortunately, most educational systems fail to 
provide what parents want and expect for their 
children. The failure is not necessarily obvious 
to the parents or fully consciously realized 
by them. They may know that something is wrong 
when they see that their children are not 
excited about going to school, and when their 
children's outlook for the future does not appear 
bright for some reason, such as low grades and
limited prospects of employment. But the parents 
may still think that the school is alright and 
the problem instead is with their children. With 
such conclusion, they see no way out of the 
situation and feel forced to accept it. 

Even if parents feel that the situation could be 
better than it is, they may not know what to do 
about it. They also do not know how to do anything
so that their children would benefit from their 
action rather than lose. For example, parents may 
be afraid of speaking up. Reasons may include a 
fear that their child will suffer consequences if 
they say too much. Also, a parent may feel 
insufficiently educated to have a say in the 
matter. 

\subsection{ What can parents do? }
\label{WhatDo}

One thing parents can certainly help with is 
providing for advanced education of teachers. 
Payment for the education of teachers can be 
arranged according to different rules than 
payments for education of children. Advanced
education of teachers does not have to be for 
free and can be supported by parents according 
to the rules they will choose to accept. 

\subsection{ Whom could parents pay? }
\label{WhomPay}

The question is whom the parents could pay for 
achieving on a regular basis that their children 
receive from their teachers the help in learning 
precisely as they need it. It is assumed that 
parents would pay experts who have a known record 
of helping teachers in this spirit. For brevity, 
we call such people experts-in-demand.

To provide some payment on a regular basis, 
parents would have to be convinced that the 
experts-in-demand whose services they pay for 
actually know and understand what kind of help 
children need and, even more importantly, that
the experts know and understand how teachers 
need to learn as adults in order to become able 
to provide the help children need. For example, 
great teachers understand that their duty is 
to help children discover and develop their 
strengths in the course of acquiring and using 
their learning skills. Teachers need to do the 
same at their level of competence, both in the
teaching as their profession and in subject 
matters they specialize in. 

Existing experts-in-demand need to be identified 
and new such experts need to be educated. Parents 
can help in this process, too, by paying for 
services of the already identified experts-in-demand. 
The experts can take care of education of their
own kind using the funds obtained from parents. 
Thus, the ultimate role of investors in the reform 
process belongs to parents and students. They, as 
the investors, seek ways to verify if their 
investment leads to expected results. Such 
verification is a duty of an investor.

\subsection{ Is learning in a curriculum? }
\label{IsLearningInC}

What the children need most is to become 
competent learners who know how to develop 
and utilize their strengths, taking advantage 
of and improving their knowledge, understanding 
and skills in agreement with fundamental values 
of human life. This goal of teaching is not 
contained in the practice of contemporary 
education of teachers as a foundation of the 
system. Teacher-proof curricula of schooling 
are not centered on the development of a student 
as a person. So, a typical teacher-preparation 
curriculum and on-the-job learning do not allow 
a misguided teacher to become a great one. To 
become a great teacher, a candidate must not 
only know the subject matter extremely well, 
which is rare, but also understand what it 
means to help students in true studies, which 
is still rarer. The amount of deliberate 
practice required for becoming a great teacher
is much greater than teachers typically have.

So, to begin a serious reform effort, even if 
only in a very small way, one needs a team of 
properly prepared experts in various disciplines.
The experts must not only know their disciplines
but also understand what it means to help a 
teacher in studying, practicing and eventually 
becoming a great teacher. These experts may 
initially collaborate only with the pre-selected 
teachers who already know that and understand 
why they need to improve their expertise in the 
disciplines in the context of which they teach 
students. In the course of this collaboration, 
teachers have a chance to learn what it means 
to have a true coach in improving one's strength 
in the process of studying, working in a team, 
producing results, and thus building one's 
system of values, knowledge and skills. Practicing 
such coaching with students teaches a teacher how
to take advantage of real-life situations as 
contexts of productive learning. Such real-life 
situations are precisely the basis of learning 
practiced by scientists, artists, and professionals.

The experts-in-demand have pressing questions of 
their own concern regarding their work with teachers. 
This means that they are involved in research on 
ways to improve teaching together with teachers. 
Such research is by no means easy. The reform 
program must develop collaboration teams including 
experts-in-demand and teachers. The results of 
work of these teams must be visible to parents in 
terms of a clear change in behavior of children 
taught by the teachers, if the parents are to 
support such work. Examples that illustrate how 
this may happen are rare. Some preliminary cases 
and hypothesis concerning mechanisms of how they 
could be expanded to larger scales will be provided 
elsewhere.

\subsection{ How much parents can pay }
\label{HowMuch}

Parents are not going to pay a lot of money 
for education of teachers in the school
where their children go to. They are also not 
going to start paying sufficiently in advance 
for educating the particular teacher who might 
later provide learning opportunities to their 
children. However, one may assume that parents 
can be prepared to pay a little bit for improvement 
of teachers in the school their kids go to, or 
will go to, if it is clear and clearly verifiable 
what they would be paying for. This assumption 
is likely to be true because people are known 
to contribute little amounts to causes they 
appreciate. For example, people certainly 
appreciate opportunities for improving life in 
the neighborhood where their life interests 
are located, and good schools are recognized 
as magnets of quality. 

In addition, the act of payment, even if the 
amount is small, is a form of a real involvement 
in school matters. It engages much more than the 
actual expenditure of money itself. It gives 
a new right to a parent as a giver. A parent 
will feel entitled to ask ``What did I pay 
for?'' Such situational design changes a private 
parent to a shareholder in the enterprise of 
educating new generations in a concrete setting. 

The actual amount will be estimated below.

\subsection{ How students can contribute }
\label{HowStudents}

Students do not have to contribute money.
Students can make the process of teaching 
infinitely easier and more productive if 
they want to learn, instead of being forced
to pretend learning and pass tests. The 
wanting is stimulated by exciting their 
curiosity and other natural features that 
students possess as extremely dynamic beings 
in rapid development. 

Senior students can help in teaching junior
students. This summit of productive learning 
and teaching methodology certainly requires 
expertise in coaching. Such coaching is familiar
in the concept of scouting and it is not clear
why such coaching is not a commonplace in 
schools. The help of students in educating 
students is probably the greatest, most 
promising and least effectively used resource 
of meaningful education. In contrast to 
practice of today, the great teachers of 
tomorrow will probably notoriously be using 
this resource with ease. 

Students can also help in running schools and 
free teachers from mundane functions that do 
not actually require involvement of advanced 
adults. Such practices of delegation of 
responsibility are also known in scouting 
organizations. By analogy with scouting, 
students can engage in learning while they do 
all kinds of helpful work that otherwise would 
have to be done by teachers and other adults. 
Consequently, if involvement of students led 
to lowering the cost of running a school, there 
would be more resources available from the 
public coffer for students, teachers, and the 
experts-in-demand. In particular, teachers 
would have more time for advanced learning 
and improvement of teaching. 

\subsection{ How parents can avoid wasting money }
\label{AvoidingWaste}

The teachers who have the opportunity to work 
with experts-in-demand would know who pays for 
it. Their advanced education through such 
collaboration should include practice in
productive communication with parents and 
allowing parents to understand what happens. 
However, parents can also judge themselves if
the funds they provide are properly used on 
the basis of observing their children. Parents 
may regularly inquire about the opinions of 
students concerning performance of teachers. 

The experts-in-demand are hired because they 
offer a high probability that as a result of
working with them the teachers will achieve 
specific changes in classroom regularities. 
The experts also propose and help organize 
activities outside classrooms. Parents can 
observe if these activities materialize in 
the improvements of school functioning using 
explicit statements made by children at home 
and, more generally, outside school, i.e., 
outside the supervision of their teachers. 
The actual rules of accountability can be 
discussed between teachers and parents as 
they decide to hire experts-in-demand.

\subsection{ Need for small trials }
\label{SmallTrials}
 
The idea of parents supporting productive 
education of teachers needs verification. 
Such verification can only be gradual, 
starting from small projects. But even a 
small project needs an estimate of its 
outlook on the future. The estimate described 
in the next section is based on the assumption 
that small projects will spontaneously create 
enough spirit of performance to form an 
{\it organization} in which students, parents, 
teachers, and experts-in-demand will be able 
to perform in a systematic way. 

Many initial attempts at small scales will 
fail. Every such failed attempt will contribute  
information about what does not work, if the 
case is recorded. If many cases are recorded, 
the information will allow participants to 
filter out elements that do work from a sea 
of inevitable mistakes. Therefore, even the 
smallest, initial projects should keep records 
of their action and results.

It is imperative not to waste resources on 
large projects that are not sufficiently 
tried out at small scales. Also, scaling up 
of every small successful project will face 
its own new problems that will require 
identification and solution. 

\section{ Calculation }
\label{sec:calculation}

The calculation described below is just an 
example that allows one to see the orders 
of magnitude that are potentially involved 
in the matter, provided that the system of 
parents supporting advanced education of 
teachers turns out to work as assumed. The 
orders of magnitude used as input data are 
realistic for Poland. In other countries, 
and for specific schools and areas that 
depart from the average values used in the 
estimate, some rescaling will be needed.
The orders of magnitude are likely to be 
the same.

A similar calculation is required in every 
lasting educational project that is supposed 
to ``do good and grow,'' if reformers want to 
assess the chance that their project can lead 
to a qualitative change over time. For example, 
the fact that the Reading Recovery program~\cite{RR} 
functions and grows suggests that more resources 
flow into the program than flow out. 

The calculation described here is carried out in 
terms of money. Poland is used as example for 
using realistic numbers. Polish z{\l}oty is about 
3 to 4 times smaller at current exchange rates 
than dollar and euro, but the average incomes and 
some basic expenses in Poland in terms of Polish 
currency are of the same order of magnitude as the 
average American and European incomes and expenses 
are in terms of dollars and euro, albeit for some
basic goods the differences are not negligible. 
Since dollar and euro are roughly the same in 
comparison to Polish z{\l}oty (1 euro $\sim$ 1.3 
\$ in 2012), the currency units are omitted. 

That the calculation is carried out using money 
is not essential. What counts is the identification 
of the source and magnitude of inflow of energy, of 
which money is treated as an indirect but useful 
measure.  

There are options mentioned in the estimate for 
significant contributions from the common tax 
money, in addition to the funds put in by parents. 
Such significant tax-based support is expected to 
eventually become a regularity as a result of 
voting by parents as citizens in a democratic 
country. Such result would indicate that the 
reform program achieves a qualitative change.

\subsection{ Example of initial assumptions }
\label{assumptions}

Suppose there is a high school that has 300 
students, see Table~\ref{tabela}. For larger 
and smaller schools, one has to properly 
rescale the numbers that follow. The number 
300 corresponds to the average size of a high 
school in Poland according to the Polish 
government statistical data from 2011. This 
corresponds to 100 students at each grade level, 
if there are 3 grades in a high school as it 
currently is in Poland. Let us assume that 
there are 100 students per grade anyway. Polish 
primary schools are on average about twice smaller 
than the average high school, and the middle 
schools are on average in between. 

On average, there are about 25 teachers per 
high school in Poland. 25 teachers per 300 
students means about 8 teachers per 100 students. 
It cannot be exactly right, but let us assume 
that 10 teachers specialize in sciences, 10 
in humanities, and 5 in mathematics. This 
assumption will yield sufficiently informative
estimates of the orders of magnitude. One can 
use these estimates for considering specific 
departures from the assumed distribution of 
specialties among teachers.

\begin{table}[b]
\caption{ \label{tabela} {\bf Numbers per high school }\\
There are on the order of 2000 high schools in Poland and 
3 times as many middle and 7 times as many primary schools.} 
\begin{tabular}{|c|c|}
\hline
students                             &  300          \\
students per grade                   &  100          \\   
teachers per school                  &   25          \\      
students per teacher                 &   12          \\                         
science      teachers                &   10          \\     
humanities   teachers                &   10          \\ 
other        teachers                &    5          \\
teachers educated/year               &    2          \\ 
students influenced/educated teacher &  150          \\ 
number of parents who pay/child      &    1  (not 2) \\       
number of units paid per month       &   10          \\
number of months                     &   12          \\ 
total units paid/year                &  36000        \\ 
units paid/teacher/year              &  18000        \\
\hline
\end{tabular} 
\end{table}

Let each of these teachers personally interact 
with 100 students every year. For example, if 
5 mathematicians teach 300 students then each 
must on average regularly interact with 60 
students. But if there are only 2 physicists in 
the faculty, each must be in regular contact with 
150 students. 100 appears a reasonable compromise 
between 60 and 150. 

Suppose that out of the 10 science teachers 
only 2 per year work with the experts-in-demand 
in an extensive program that is proven useful 
for the purpose or reforming education with
parents involved in covering the cost of such 
collaboration (see below).

The number of teachers in the program is 
assumed to be only 2 per school because 
the program is expensive and the smaller 
the number the smaller the burden for 
parents. But the number of teachers per 
year cannot be smaller than 2 per school 
because they need to support each other in 
solving problems they encounter in their 
school.

The collaboration with experts-in-demand 
does not take a teacher out of her or his 
school. Instead, the experts-in-demand 
provide help on-site or in a nearby location 
to which the 2 teachers can easily commute 
within a working day at least once a week. 

The two teachers who receive this special 
opportunity are assumed to understand and 
appreciate it that they belong to the team 
of faculty in their school and that their 
professional goal is to share what they 
learn while they serve the students in their 
school and communicate with other teachers. 
These teachers are assumed to want to grow 
personally as professionals in their trade.
 
Assume that more teachers in the school 
rotate over years as collaborators with 
experts-in-demand at increasing levels of 
sophistication. The better and more evident 
the results of such collaboration year after 
year the more teachers of the same faculty 
team want to participate in working with the 
same group of experts-in-demand. So, assume 
that the 2 trained teachers share what they 
learn among all 10 science teachers in their 
school. Their new experience includes ideas 
and practice of how to share what they learn 
with faculty members in their school. These 
assumptions mean that the advanced education
of one teacher a year influences education
of more than 100 students a year because 
teachers learn from each other on the job 
and mutually improve their teaching. 

In order to make the order-of-magnitude 
estimates, one can assume that advanced
studies by one science teacher a year 
serves improvement of education of 150 
students a year. If the advanced study
program indeed improves the performance 
of teachers, the whole school benefits. 

\subsection{ Informing parents }
\label{informing}

Informing parents about the changes that 
occur in school practice as a result of 
the advanced studies, especially when the 
changes touch the roots of school culture, 
is very important locally. Informed parents 
have a chance to understand that if they 
contribute to the advanced studies of just 
two teachers a year then the whole school, 
including their own children, benefits in 
many ways. The indicators of success and 
profitability of investment that parents 
can monitor include: changes in the behavior 
of children, content and form of communication 
between parents and teachers, and quality of 
activities that children participate in and 
outside school and describe to parents at 
home. One of the perceptible effects is the 
excitement about future that results from 
concrete events, such as a completion of an 
inspiring task by a group of students, a 
result worth thinking and talking about.
For example, as far as teaching in the context
of science is concerned, students may go on
a research trip arranged according to the 
plan that they and their teachers design in 
collaboration with experts-in-demand.

\subsection{ Cost per teacher per year }
\label{costperteacher}

The chain of interdependent factors described 
in previous sections leads to the assumption 
that parents of 150 students (half of our 
model school population) may cover the cost of 
advanced studies of one teacher per year in
collaboration with experts-in-demand. Parents 
of one student may pay just a little bit. All 
parents of all students in our model school 
(300 students) may thus support advanced 
education of two teachers per year.

How much could a father or a mother pay per 
year for the purpose of improving the teaching 
standards of the school their child goes to? 
If they paid just 1 unit of local currency per 
month, they would pay 12 units a year. If they 
paid 10 units per month, they would pay 120 
units per year. When the latter is multiplied 
by 300 (the number of students in the school), 
the budget for advanced studies by two teachers 
per year turns out to be 36 thousand, which 
means 18 thousand per teacher per year. This 
amount is comparable to tuitions in respectable 
public colleges. 

The parental support means that the regular 
school budget remains untouched despite that 
new quality is flowing in into the school 
operation. The increase in quality is correlated
with the parental involvement, transparency of 
funding for advanced learning of teachers, and 
confidence of purpose that fuels the spirit of 
performance in teaching students for the 
entire school faculty. 

Some parents may not like the idea of paying 
even if it is only 10 units of currency per 
month. But some may be ready to pay more. The 
fact that the issues of payment for improvement 
of teaching quality may be discussed and acted 
upon is here assumed to have positive consequences 
for the school community. 

In any case, the assumptions made above amount
to one parent per child paying 10 currency 
units per month. One might also assume in 
many cases that two parents could pay such 
amount per child. This would result in the 
increase of estimated funds by factor 2. 
Thus, assuming only one parent per child 
pays may be underestimating the inflow of 
resources that parents may be willing to
provide on regular basis. In the future, 
when they are convinced by the long-term 
record of concrete results, parents may 
not even think about questioning that it 
makes sense to provide support for advanced 
studies by teachers in collaboration with 
experts-in-demand.

\subsection{ Where does the tuition go }
\label{tuition}

Suppose the tuition for advanced education 
of a teacher goes from parents of students 
in a school directly to a special company. 
Let us call this company a Teachers Center 
(or TC). The TC is formed by the experts-in-demand. 
One can think about TC in terms of an analogy with 
a medical research group formed by all kinds
of experts who unite in effort of helping 
doctors improve their services to patients.

\subsection{ Comment concerning the goal of TC }
\label{TCgoal}

It is essential that the advanced studies by 
teachers in collaboration with the experts-in-demand 
who form a TC has a clear, overarching goal. 
{\it The central task of a Teachers Center is to 
build, use, improve, and share the concepts of 
productive learning and productive teaching among 
teachers and professionals and through them and 
their students in the whole society.} Unless a TC 
performs this task, it loses sense as an enterprise. 
However, it is not relevant what its central task 
is for the purpose of just crunching the numbers 
and estimating the resources that parents of 
students in schools in the neighborhood may 
actually provide for supporting the TC that 
functions well.

The due explanation of advanced studies process 
for teachers is a difficult task that requires 
expertise. The key difficulty is to explain the 
concepts of productive learning and productive 
teaching. The explanation cannot be provided or 
comprehended using only words. Reading is not 
enough. A long-term commitment to study of theory 
and practice of learning and teaching in various 
contexts is required of a person who wants to 
understand what is involved. 

The concepts of productive learning and teaching 
are not commonly recognized as such precisely 
because the society's concepts of education are 
built through experiences within the current 
system. This system is based on the idea that 
human minds can be filled with information at 
school like books that are printed in a printing 
shop. The idea of mind-printing is about four 
centuries old. It dates back to the invention 
of print and educational system design described
by Comenius. Today, the printed book is being 
replaced by interactive technologies. With their 
help, humans can emancipate their learning from 
the regime of teaching based on the concept of 
printing. But it does not mean that the concepts 
of productive learning and teaching that are 
essentially interactive become broadly recognized 
right away as a result of emergence of new 
technologies. In the case of printing, nearly 
two centuries passed between the invention in 
technology (Gutenberg around 1450) and a conceptual 
design of a new educational system that works in 
analogous ways (Comenius around 1650).

The lack of shared concepts of productive learning 
and teaching is the actual reason for this article. 
Namely, the estimate is provided for the purpose 
of indicating that a continuous, interactive reform 
process, capable of leading eventually to a situation 
in which a short document on the principles of 
productive learning and teaching could be considered 
clear by a majority of readers educated by a universal 
public system, can be initiated by teachers, sustained 
by parents, and helped by experts-in-demand.

\subsection{ Crunching numbers for a TC}
\label{TCnumbers}

Let a TC work with about 100 teachers at a time. 
Two teachers per school means that a TC serves 
50 schools. In Polish realities, where there are 
16 provinces (16 wojew\'odztw) and on average 
125 high schools per province, this means that 
there are 2 to 3 such parent-supported TCs per 
province. If all high schools in Poland were 
served and every TC served 50 schools, one would 
have about 40 TCs in the country. Our estimate 
calculation begins with these numbers (see 
Table \ref{tabelaTC}).

The TC budget that originates directly from 
the pockets of parents is estimated as 100 
times 18 thousand, which amounts to about 
1,800,000 a year. It is thus not entirely 
out of order to assume that a TC can operate 
in a stable fashion. 
\begin{table}[b]
\caption{ \label{tabelaTC} 
{\bf Estimates for Teachers Centers (TC).} } 
\begin{tabular}{|c|c|}
\hline
number of teachers served/TC/year           &  100            \\
number of schools served/year               &   50            \\ 
number of TCs per province                  &  2-3            \\
tuition collectable per TC/year             &  1.8 mln        \\
budget including matching funds/year        &ca 3-5 mln     \\ 
number of students served/TC/year           &  15000          \\
number of great teachers/TC/10 years        &  $\gtrsim$ 100  \\
number of great teachers/all TCs/10 years   &  4000           \\
\% of great high school teachers            &                 \\
after 10 years/Poland                       &  8\%            \\
\% of great high school teachers            &                 \\ 
after 30 years/Poland                       &  24\%           \\
number of staff members/TC                  &  18             \\   
units inflow per staff member (conserv.)    &  $\sim$100,000  \\
units inflow per staff member (optimist.)   &  $\sim$300,000  \\
number of schools served/employee           &  16-17          \\
number of schools served/day/employee       &  3-4            \\
number of teachers served/day/employee      &  6-8            \\ 
\hline
\end{tabular} 
\end{table}

A TC that is widely recognized for its 
verified utility to teachers and thus 
students in schools, is expected to be
capable of obtaining a share of tax money. 
There can be matching funds from the local 
government. There can also be matching 
funds from the central government in 
support of the local government for this 
purpose. 

In addition, it is also possible that private 
donors contribute. This possibility includes  
not only the local citizens and employers, 
but also outside investors and buyers who 
are interested in patterns and sources of 
educational solutions that apply on a much 
wider scale than only locally. A TC may publish
and sell its expertise for a hefty fee.

If the additional sources contribute to the
inflow of support, the total TC budget may 
climb to twice or even thrice the amount 
provided by local parents alone, i.e., it
may grow to 3,600,000 or even 5,400,000. 
This does not include yet professional grants 
that may be awarded to a TC for specific 
projects.

\subsection{ TC productivity }
\label{TCproduct}

A TC that serves a community of 50 schools 
of 300 students each serves 15,000 students 
a year. After 10 years, the TC could have 
served 1000 year-teachers, most likely many 
of the teachers being involved in a year of
advanced collaboration with experts-in-demand 
more than once, i.e., in a sequence of gradually 
increasing levels of competence in learning 
and teaching. As a result, the TC appears to 
have a chance of producing at least 100 great 
teachers in 10 years. The number 100 is 
obtained if one and the same group of 100 
teachers collaborates with a TC over the whole 
period of 10 years. If more teachers are 
involved, the resulting number of great 
teachers may be greater than 100 per 10 years.

If there are only a bit more than twice the 
number of TCs than provinces in Poland, say 
40, after 10 years of operation there could 
be at least 4000 great teachers in high schools 
around the country, including and in addition 
to the ones who already work there. Since there 
are about 50 thousand teachers in high schools 
in Poland, the 4000 would amount to about 8\% 
of all teachers. Thus, one can estimate that 
after 30 years at least about a quarter of all 
high school teachers would be the great ones, 
fully conscious of the value of their services 
to society and capable of expressing their 
competent opinions concerning educational 
reforms. These teachers would also be capable 
of forming a strong professional organization 
of the kind that teachers cannot currently form, 
even if they are great. Such strong, self-conscious
professional organization could double its 
membership faster than in 10 years.

\subsection{ Funds per TC employee }
\label{perTCemployee}

Suppose a TC has a budget of only 1,800,000 
currency units, which means that there are 
no matching funds, donations, sales, or grants. 
How many staff members could a TC have? If it 
had only 18 permanent staff members, such as 
3 great teachers of whom one is the TC director,
5 mathematicians, 3 physicists, 3 chemists, 3 
biologists, and 1 properly prepared person
for administration, there would be 100,000 
units of currency flowing in per year per 
employee to cover their salaries, benefits, 
and overhead. This is a conservative estimate.

If there is a few times larger budget, the 
numbers change. In a well-performing TC, there 
may be more permanent staff members, they may 
earn more per hour, and there may be funds for 
more equipment and travel for study. The factor 
3 being not unthinkable, as indicated above 
assuming tax money, donations, sales, and grants, 
one can consider an optimistic estimate of 300 
thousand a year of TC funds available per 
employee in the area of science.

Some of the numbers would double in an estimate 
including humanities. The doubling would require 
20 units paid per month by one parent per child, 
or 10 units paid by two parents each per child. 
The essential benefit in the case of Poland would 
be, however, that in 30 years the great teachers 
in Polish high schools would form about 50\% of 
all high school teachers. The period of 30 years 
is not even 10\% of the time over which the 
current system developed since Comenius offered 
his design. 

If the TC system worked also for primary and
middle schools, it would form a major force
of change in the hands of parents, teachers,
and experts. 

\subsection{ TC workload estimate }
\label{TCwork}

Can 3 experts-in-demand who specialize in 
physics serve 50 schools productively? One 
of them would have to serve about 16-17 
schools. This means 3 to 4 schools a day 
during 5 working days of the week. But if 
in each school only 2 teachers are 
participating, there are only about 6 to 
8 people to interact with per expert-in-demand 
per working day. Still, every one of these 
experts would have to have extraordinary 
knowledge and skills in order to perform 
the required tasks of assisting and coaching 
6 to 8 teachers a day. 

In the demanding scenario of 6-8 teachers 
per expert-in-demand per day, there would of 
course emerge problems of logistics. But the 
schools may cooperate. On one day one group
of 6-8 teachers may meet in one school, and 
on another day another group may meet in 
another school. 

On the one hand, the logistics may appear as 
a big problem. On the other hand, it provides
a great opportunity for the teachers of one 
school to learn about what happens in other 
schools. Teachers may develop their inventiveness 
and entrepreneurial attitudes in a collaboration 
that requires solving problems of logistics. 
The contacts necessary for solving the logistics 
problems may evolve in time into voluntary 
engagements and commitments that accelerate 
sharing of the concepts of productive learning 
and teaching, rooted in real-life contexts. 

\section{ Formation of a TC }
\label{TCformation}

Since the experts-in-demand are hard to 
find, the staff of a TC would have to 
be completed in a careful process of 
selection. The selection starts with 
identifying the first director, an actual 
founder of a new TC. 

\subsection{ The TC director }
\label{d}

The director would have to be a great teacher 
who deeply understands the concepts of productive 
learning and teaching in theory and practice, is 
recognized as a great teacher by other teachers, 
qualifies in terms of competence and integrity as 
a manager and a leader, wants to create the 
profession of teaching, and is prepared for the 
creation of a working TC to take a lot of effort 
and time. 

The director's most important first task 
include identifying the teachers who want 
to form the TC and finding the expert- or 
experts-in-demand who can provide the initial
help the director knows to be truly needed by 
the teachers in her or his school. Personal 
contacts between the director, teachers and 
experts-in-demand will form the foundation 
for further steps because nearly all these 
steps require personal safety and trust. The 
first tasks also include convincing the initial 
group of teachers in a school that inviting 
some expert-in-demand can help in solving the 
problems that the teachers face. The group 
of convinced teachers can subsequently appeal 
to parents for providing the minimal necessary 
amount of funds that are sufficient to cover 
the cost of first visit or visits by the 
identified expert or experts. 

Another mandatory task in the initial activity 
is to keep a record of events and their results, 
including results of the teachers-experts 
collaboration. The record will be a key to 
demonstrate in terms of description of facts 
that the teachers-experts collaboration is 
productive. For example, one can record new 
elements in teacher-student conversations, new 
features in the behavior of students, new 
decisions of teachers regarding their classroom 
work, and reports of parents on changes in 
behavior of students outside school. 

\subsection{ After-school laboratory }
\label{aflab}

The director must respect systemic constraints on 
the school functioning. Thus, opportunities for 
experimentation by a teacher and an expert-in-demand 
exist in the form of an after-school program that 
they can design, create and run together with students. 
The after-school program is a source of new ideas 
about what and how can be improved in working with 
students, also in the regular classroom. A separate 
document will discuss the key role that a school-based 
scouting organization can play in this context. 

Since this note is focused on the estimate of
cost of reform to parents, the key question of
how scouting differs from Comenian-like school 
based on analogy with printing is not discussed 
here in any detail, except for a few comments 
below. In particular, it is essential to avoid 
the impression that the after-school educational 
laboratory programs that draw on the concept of
scouting can be thought about as childish. Three 
examples are invoked here to stress how important 
the scouting principles are. They concern fundamental 
issues of the role of productive learning and 
teaching in functioning of democracy and improving 
the quality of life of people.

The first example to mention here concerns the 
tradition of Polish scouting, documented in many 
publications of which only three are quoted here~\cite{Wachowicz,Janowski,Kaminski}. These 
books are available in Polish. Their existence 
only in Polish illustrates the need for creating 
an international organization that is capable of 
employing educational resources available from 
different peoples in different languages. The 
books make it clear that the concepts of scouting 
adopted in Polish culture in 20th century on the 
basis of struggle for freedom and democracy at 
least since 18th century led to the creation of a 
great organization that provided Polish youth with 
values and wisdom they later used as adults, recently 
in overcoming communism and during most recent 
decades in developing a free society.

The second example is provided by the Girl Scouts 
of the USA, essentially captured by Frances 
Hesselbein in her book {\it My Life in Leadership}~\cite{Hesselbein}. 
Hesselbein demonstrates the seriousness of issues 
that scouting is concerned with in the US. She 
also describes the examples of recognition of 
values of scouting principles and educational 
methods among leaders of the American people. 
This recognition exhibits itself in the existence 
of Leadership Institute~\cite{LeadershipInstitute}. 

The third, fundamental reference concerning the 
laboratory for advanced studies by teachers 
according to our estimate, is provided by Peter 
F. Drucker~\cite{Drucker}. In a discussion with 
Albert Shanker, former president of the American 
Federation of Teachers, Drucker stresses the 
importance of foundation of achievement, which 
requires high-quality education needed by independent 
knowledge workers. Shanker points out how productively 
he learned in a process of earning merit badges 
when he was in Boy Scouts. The point concerning 
the estimate described here is that parents can 
help in joining the two ideas: one of the required 
competence that is stressed by Drucker and the 
other one of how one actually achieves competence 
that is stressed by Shanker, by supporting 
creation and functioning of TCs in which teachers 
can study the variety of issues that matter in
teaching a person and where they can interact 
with students in ways akin to best tradition 
of leadership in scouting.

An after-school program allows the teacher to 
develop and experiment with her or his ideas 
and apply them as they develop in collaboration 
with the expert-in-demand and students. The key 
aspect of the after-school program in this context 
is that the teacher and the expert-in-demand 
can engage in the process of advanced learning 
about students in ways that are free from pressure 
of the existing educational system. In other words, 
they can focus on questions concerning the unknown, 
the unclear, the difficult, the never-understood, 
and the always-suppressed because of all kinds of 
fears. 

In particular, the fears stem from the low self-esteem 
and uncertainty about personal status, especially 
when a teacher is aware of loopholes in her or his
comprehension of the world and may easily be accused 
of lack of familiarity with the depth of knowledge 
and degree of understanding that are necessary for 
being a teacher. One's comprehension of the world 
and familiarity with knowledge and understanding 
of it that is currently available to humanity is 
the actual basis of what a teacher can truly do to 
help her or his students in becoming learners. 
The informal after-school laboratory allows all 
participants to ask the questions that they are 
afraid to ask in the formal system setting where 
they can be unfavorably judged if they disclose 
blanks in their knowledge, loopholes in understanding, 
or shortcomings in skill. 

The after-school format allows students to pursue 
their interests in forms that are not acceptable 
in the current educational systems, such as field 
studies over extended periods of time far away from 
the school building and according to the wishes 
and ideas of students rather than teachers who 
must comply with the curriculum and schedule imposed 
from above. A teacher who is capable of earning the 
trust of students to the extent that allows her or 
him to naturally participate and observe students 
in such activities, is extraordinary rare in the 
current systems. In the reform program that exploits 
the opportunities offered by the after-school programs 
which take advantage of learning principles employed 
in scouting, such teachers would be the leaders.
 
\subsection{ Initiative in the hands of teachers }
\label{initiative}

The initial group of teachers may proceed with 
increasing the number of hours of interaction 
with experts-in-demand provided that they are 
funded by parents on the basis of the results 
that teachers obtain with students, results
clearly demonstrated to parents. Thus the process 
of formation of a TC is essentially based on the 
will and competence of the founding teachers. 
It is never imposed on them in any form from 
above. 

Unless the initiative comes from the depths 
of hearts and minds of teachers and is met 
with competence of experts-in-demand so that 
the results of their collaboration are clearly 
recorded, the TC will not create a solid, 
documented base for its long-term development 
with support from parents and thus will not 
provide the full impetus it could to any
wider reform effort.

The teachers have to consciously tackle 
the problem of identifying experts who 
can understand the goal of forming a TC 
sufficiently well for making a reliable 
commitment to helping founders of a TC in 
solving subject-matter problems that 
require advanced studies. For example, 
the founders of a science TC would have 
to arrange voluntary agreements with 
mathematicians, physicists, chemists, 
biologists, etc., who would be able to 
provide help of the kind needed by great 
teachers and by candidates for great teachers. 
This includes help in learning about the 
mechanisms of progress in the disciplines 
that teachers study. These mechanisms
typically involve a host of features that 
are essential in productive learning and 
teaching. 

Undoubtedly, the group of teachers would
immediately face many foreseeable and 
unforeseeable problems. Their chance for 
success would depend on the number of 
teachers in the founding group and the 
strength of their desire for growth and 
advanced learning in collaboration with 
scientists, artists and professionals
who collaborate with then as the experts-in-demand. 

\subsection{ Pedagogy and andragogy}
\label{PandA}

The teacher founders of a TC would have to be 
in touch with the experts on pedagogy and 
andragogy (one can learn about the latter 
following attempts of Malcolm Knowles). These 
experts would have to be competent enough to 
advise both the teachers and experts-in-demand 
on the issues that emerge on the way toward 
productive functioning of a TC. 

Voluntary engagements in the area of pedagogy 
and andragogy could be supplemented with special 
projects that obtain funding in the form of 
grants for definite research purposes. Such 
research would be of interest to professionals.
Therefore, the professionals would want to carry 
out research concerning the formation and functioning 
of a TC. This subject may actually become a new 
area of research in pedagogy, andragogy, and many 
other disciplines in social sciences anyway.

The service of professionals to a TC could 
be recognized in their home institutions as a 
part of the duties that professionals are expected 
to fulfill as members of society. Such services 
build the status of social responsibility for 
their home institutions.

\subsection{ Self-selection of founders }
\label{self-selection}

Most importantly, however, a formation of 
a TC needs founders who can function as 
leaders. Such founders can currently only 
self-select. This note is supposed to help 
the potential leaders to self-select. 

A natural step of self-selection for a founder 
of a TC would be to contact the author with 
a suggestion of what step or steps the founder 
would like to make toward creation of a TC. 
A realistic suggestion could provide a basis 
for discussion on how to go about making these
steps.  

The above comments apply to the stage of 
creating first TCs. The ones that survive 
the initial phase and grow would become 
sources of insight concerning how to form 
new ones.

\subsection{ TC location and operation }
\label{TClocation}

Where would the TC be located? This is 
particularly interesting and important 
question for this note. The old paradigm 
of a center is a building where people 
assemble for some purpose, like in a 
biggest hut in a village. Today, 
electromagnetic and electronic means of 
telecommunication do not force us to think 
in the same terms. Thus, while a TC should 
be located somewhere in the middle of the 
50 schools it serves for reasons to be 
explained below, the concept of the middle 
refers not merely to geography but to the 
center in the world of values, knowledge 
and skills that are relevant to the purpose 
of serving the school teachers whose students' 
parents support the TC.

The basic reason for geographic location 
of a TC in the middle of the area where 
the teachers it serves work and live is 
that such location makes it easy for the 
teachers to meet in the TC. However, one 
can also consider the concept of a virtual 
TC. Such virtual TC could secure easiness 
of personal interaction among teachers and 
experts-in-demand using the school premises 
where the teachers work. These buildings 
already exist and they are not fully utilized 
outside the times specified in the regular 
classroom schedules. 

Currently, however, the function of a central 
hut in a village cannot be treated as marginal. 
For example, church buildings are erected for 
the purpose of assembling and uniting by 
like-minded people. The fact that a church 
building exists is providing people with a 
sense of stability of foundation of their 
conceptual frame of reference. It visibly 
confirms the commitment of the attending 
community to sustain the organizing function 
of this frame of reference.

Can a TC be built? If it is so decided by a 
sufficiently large number of teachers in 
sufficiently many schools, they may undertake 
efforts to secure the necessary resources. 
Parents are expected to help. Local governments 
may help with land, construction, and operation 
cost. After all, the teachers who are to use 
the building would teach the local kids. If 
the parents of these kids want that a TC has 
a building, the local governments can be 
expected to support their wish with funding, 
since otherwise they risk losing election. 

On the other hand, a virtual TC may exist as 
an organization of teachers and experts-in-demand 
who do not need an entire building. Besides taking 
advantage of school facilities that already exist
and can be used after school hours as necessary, 
teachers and experts-in-demand may also take 
advantage of telecommunication technologies (as 
they develop over time). In particular, the mode 
of working from their residences using efficient 
means of telecommunication is often the preferred 
mode of functioning by creators and independent 
thinkers. They can utilize cutting edge 
telecommunication technologies precisely as far
as they can be helpful. 

\subsection{ Promise of the future }
\label{promise}

The estimate described above provides the promise 
of the future to all stakeholders who engage in a 
project whose goal is formation, functioning, and 
continuous improvement of a TC as an element in an 
entire network of the TCs. Namely, for as long as 
there will be parents and students, the estimate 
says, the future does not have to be questioned if 
the TCs serve their mission well. 

Whether the TCs do or do not perform as needed is 
continuously verified by parents on the basis of 
how their children perform. This direct accountability 
scheme prevents the network from developing disparities 
and opportunities for abuse. The participants can react 
to indications of problems long before a pile of problems 
leads to a major malfunction.

\subsection{ Initial action }
\label{firstaction}

What every teacher vitally interested in 
authentic educational change can do already 
next Monday is to start talking with other 
teachers and people of similar interests about 
how to begin building a local organization 
that will be able to identify and hire first 
experts-in-demand for working with local 
teachers on their problems. In the practice 
of a regular school, this means that teachers 
find a way to talk with parents and ask if the 
parents could support a visit of some expert. 

A good way to talk appears to be to first
demonstrate to parents on a small example 
the kind of benefits that such visit may 
bring. This is a subject for a thick book. 
Then, a carefully selected expert-in-demand 
can be proposed for invitation for a definite 
purpose that parents will understand. Since 
the cost is to be covered by parents, the 
expert-in-demand should be invited for the 
purpose of solving a particular problem which 
the teachers truly want to solve and about 
which the parents know it is acute, urgent, 
and not easy to solve. 

For example, the problem may be that parents 
are poor and have no funds to support an 
excursion of students to a major place of 
interest, such as a science museum in a 
capital. If children earn a portion of the 
money that they need for the trip by doing 
useful work in their community, their thinking 
about money will change. Parents will certainly 
notice that their cost of providing for education 
of their children is lowered in an unexpected 
way. If the idea of proceeding this way and how
to do it comes from an expert-in-demand, 
materializes in the hands of teachers, and 
parents know the results, they may be convinced 
by the group of founding teachers that it makes 
sense to provide resources needed to invite the 
expert-in-demand whom the teachers want to 
come to their school for a well-defined reason
relevant to creation of a TC. The cost per 
parent may be small. Large direct monetary 
expenses can be avoided. Instead of paying for 
a room in an expensive hotel some parent may 
host the guest. Food may also be arranged not 
necessarily in an expensive restaurant, etc. 
If the expert lives nearby, the cost may be 
entirely negligible in comparison to the 
value of outcomes.

Another example of a serious problem is 
flooding, apparently local. Typically, 
the problem of flooding cannot be solved 
locally and requires a broad comprehension 
of its origin in order to figure out ways 
for seeking a solution. In the course of 
learning how to deal with such serious problem, 
an expert-in-demand can play the role of an 
advisor to the teachers and parents about how 
to help children comprehend what actually 
happens, what needs to be done, and how to 
do it. The excitement about and desire for 
finding solutions may greatly influence how 
children learn and how they think about their 
future. The change will be visible to parents 
in terms of conversations the children will 
initiate with them at home and the conclusions 
children will openly come to regarding vital 
issues.

Still another example of a serious problem 
concerns informing parents about the difficulty
of subjects of major interest that teachers are
supposed to explain to students, such as what 
the place of humanity is in the universe, where 
nuclear power is coming from, how it was 
discovered, and how it can be used or misused,
what we know about human brain and its development 
in evolution and in life of one person as it learns, 
etc. The problem is that the school has a mission 
of helping students learn about the world where 
they live. Parents may not fully realize the 
challenges that teachers face. A suitable program 
designed by teachers with participation of 
experts-in-demand may help parents recognize 
that teachers need competent help in teaching 
students. One way to explain this challenge to
parents is to ask an expert-in-demand to come
to a meeting with teachers and parents at their
school and talk about it in the form of a lecture, 
a discussion, or some other activity that the 
teachers and experts can invent.

If a series of visits by experts with support 
of parents is successful, teachers may eventually 
identify experts worth an extended collaboration 
on issues of productive learning and teaching. 
This activity would allow teachers to obtain the 
initial evidence needed for appealing to parents 
for a sustained support and possibly a contribution 
to creation of a precursor of a future TC for a 
number of schools. The latter step would obviously
require an intense, self-motivated collaboration 
of teachers from more than one school, a seldom 
found phenomenon in today's systems.

\section{ TC network needs a brain }
\label{NTCbrain}

Where could the assurance of wisdom and quality 
of a TC as a seed of change supported by parents 
come from? The point of the estimate described 
in this short note is not to explain the full 
answer to this fundamental question. The answer 
is only hinted at below after the following 
remark.

The purpose of estimate described in this note 
is to say that gradual creation of a new 
system does not appear hopeless, even if it is 
not easy and must take time. Namely, the estimate 
says that there is enough energy available in 
parents and students to create and sustain a 
whole Network of Teacher Centers (NTC). Employees 
of the NTC would work as scientists, artists, and 
other professionals always do, except that their 
focus would include productive learning and teaching. 
There is no point in discussing how they would 
actually do it. This cannot be explained in words
on a page or two. The readers interested in the 
issue will have found required elements already 
described in relevant literature.

However, there is one feature of the NTC that 
must be stressed immediately: {\it the NTC needs 
a brain.} 

Mother Nature tells us that every advanced 
living organism has a brain. Similarly, the 
TCs cannot function in isolation from each 
other and from sources of power that drive 
society. The strength of their function must 
come from integration in a greater culture. 
The NTC needs a brain that enables it to 
constantly earn and hence maintain the 
society's support. 

{\it The predictably dominant function of the 
NTC brain is to maintain, continuously improve, 
and guard the process of preparing great 
teachers, i.e., professionals who understand 
and excel in utilizing and improving the concepts 
of productive learning and teaching for the 
purpose of helping people become, generation 
after generation at all ages and levels of 
competence. The NTC brain constantly learns
how to best perform its function.}

An example of a precursor of a required structure 
and function appears to exist in the form of 
Reading Recovery (RR), whose creation was greatly 
advanced by Marie Clay, an educational expert par 
excellence~\cite{RR}. Marie Clay has carried out 
her research on difficulties children encounter 
when learning to read for nearly two decades in 
New Zealand. Later, the RR program matured enough 
to encourage people to implement it on a broader 
basis, far beyond New Zealand. RR operates now in 
the US and Europe. It was born having a learning
brain. 

There are at least two reasons for the NTC 
brain to be international. One is that no 
country has enough top expertise on every 
issue that the preparation of teachers involves. 
Another reason is that if the brain is not 
international, the organization will be prone 
to develop nationalistic attitudes that lead 
to disasters, such as wars or large, strong 
nations exploiting weaker, less educated and 
less resourceful ones.

Formation of a potent brain in the head of a 
meaningful reform program for education is 
the challenge of highest intellectual caliber 
imaginable, at all levels. The challenge is 
by no means limited to educational institutions 
because it extends to major professional 
organizations from which the experts-in-demand 
originate. 

\section{ Conclusion }
\label{sec:c} 

The principle of parents helping to take care 
of educational reform by supporting advanced 
learning by teachers has apparently never 
been consciously applied in educational reform 
on a global scale. Nothing is known to the 
author about its feasibility on such scale. 
The global system may only emerge from trials 
and collaboration of many small projects and 
only if they succeed in finding ways and means 
for a broad collaboration. The hope is certainly
associated with students who receive the new
kind of preparation and take over the duties
of great teachers and experts-in-demand in
carrying the process of change on from 
generation to generation.

Small or large, independent educational projects 
will not necessarily be welcome. It is likely 
that creation of NTC supported by parents will 
be opposed by people who lack the required 
insight. For example, the excessive bureaucracy 
in systems where parents have no real role to 
play is likely to consider the idea as threatening 
and counterproductive. The bureaucracy cannot be 
blamed for thinking this way since they were not 
involved in deciding what kind of an educational 
system molded their minds. But their deeply rooted 
resistance must be taken into account in planning.

The principle of parental support for advanced
studies by teachers is also likely to be 
rejected by those who see themselves as capable 
of and thus {\it entitled} to taking advantage 
of resources of our world without much of a 
thought about where the resources they use 
actually come from and without acknowledging 
that they use these resources at the expense of 
others who are less lucky and do not realize what 
happens. For example, this is how members of 
strictly profit-oriented and corrupt organizations 
may be expected to respond. 

Organizations that exploit others by operating 
from their privileged position and taking 
advantage of ignorance of the exploited will 
find the concept of NTC particularly dangerous. 
Namely, some parents may recognize and understand 
when youth is being exploited. If these parents 
secure through their support of contacts between 
teachers and independent thinkers that the teachers 
learn what happens and explain it to students, 
the exploiters will not be able to continue their 
practices of taking advantage of the ignorance 
and naivety of youth. 

One also needs to prepare for the difficulties 
associated with creative destruction that a
new approach always faces in the process of 
replacing less productive approaches. Practitioners 
of the latter will fight for survival. The fight 
for survival means that they will use any means 
they can to avoid defeat in competition for the
parental support.

The optimistic aspect of the estimate described
in this note is that it does not require everybody 
to perfectly understand and agree on everything in 
advance. People can start supporting the process 
of change by small steps, such as just one school 
using funds provided by parents to hire one or 
two experts to help just one team of teachers, 
etc. The estimated costs of the small startups 
to parents appear surprisingly low in comparison 
to the value they are capable of providing.

\newpage

{\bf Acknowledgment }\\

Among many people who influenced his thinking, 
the author gratefully acknowledges many illuminating 
discussions with Seymour Sarason concerning the 
nature of human learning and formation of settings
(Seymour died in 2010). The author also greatly 
benefited from multiple discussions with Kenneth 
Wilson concerning learning and systemic issues in
the context of physical sciences, Reading Recovery, 
and the conceptual legacy of Peter Drucker. In 
addition, the author thanks Andrzej Janowski for 
his work on creating and running the youth organization 
called 1 WDH ``Czarna Jedynka'' that influenced the 
author in his youth (and for posing stimulating 
questions concerning the lessons drawn by the author 
from the concepts and practices of that organization 
much later).

\newpage
 
\begin{table}[t]
\caption{ \label{tabelaPoland} 
{\bf Statistics of Polish 
     schools in 2011 \cite{GUS}.} } 
\begin{tabular}{|c|r|}
\hline
primary schools (6-13)        &   14000  \\
middle schools (13-16)        &    7000  \\ 
high schools   (16-19)        &    2000  \\ 
colleges and universities     &     400  \\ 
high schools/province         &     125  \\
teachers of primary schools   &  176000  \\
teachers of middle schools    &  109000  \\
teachers of high schools      &   50000  \\
students of primary schools   & 2200000  \\
students of middle schools    & 1300000  \\
students of high schools      &  630000  \\
students/primary school       &     160  \\
students/middle school        &     186  \\
students/high school          &     315  \\        
students/teacher in primary   &      15  \\
students/teacher in middle    &      13  \\
students/teacher in high      &      12  \\
teachers/primary school       &      13  \\
teachers/middle school        &      16  \\
teachers/high school          &      25  \\
average teacher income/month  &    3400  \\
teacher work cost/hr&      42  \\
estimated number of hrs/month &      80  \\       
\hline
\end{tabular} 
\end{table}

\end{document}